\documentstyle[psbox]{WORKSHOP}

\begin{document}

\title{MAXI\,J1659-152: the shortest orbital period black-hole binary}

% AUTHOR(S) 
\author{
E.\ Kuulkers,$^1$ C.\ Kouveliotou,$^2$ A.J.\ van der Horst,$^3$ T.\ Belloni,$^4$ J.\ Chenevez,$^5$ A.\ Ibarra,$^1$ \\
T.\ Mu\~noz-Darias,$^4$ A.\ Bazzano,$^6$ M.\ Cadolle Bel,$^1$ G.\ De Cesare,$^6$ M.\ D\'\i az Trigo,$^7$ E.\ Jourdain,$^8$ \\
P.\ Lubi\'nski,$^9$ L.\ Natalucci,$^6$ J.-U.\ Ness,$^1$ A.\ Parmar,$^1$ A.M.T.\ Pollock,$^1$ J.\ Rodriguez,$^{10}$ \\
J.-P.\ Roques,$^8$ C.\ S\'anchez-Fern\'andez,$^1$ P.\ Ubertini$^6$ and C.\ Winkler$^{11}$
\\[12pt]  % TO BE SPACED WITH ONE LINE
%
% INSTITUTES OF AUTHORS
$^1$  European Space Astronomy Centre, SRE-O, Villanueva de la Ca\~nada (Madrid), Spain \\
$^2$  NASA/MSFC, USA $^3$  USRA, USA $^4$ INAF - Brera Observatory, Italy \\
$^5$  DTU Space, Copenhagen, Denmark $^6$  INAF/IASF Rome, Italy $^7$  ESO, Garching, Germany \\
$^8$  IRAP, Toulouse, France $^9$  NCAC, Toru\'n, Poland $^{10}$  CEA, Saclay, France \\
$^{11}$  ESA/ESTEC, The Netherlands\\
%
% please put the first author's initial and e-mail address below
{\it E-mail(EK): Erik.Kuulkers@esa.int} 
}

\abst{Following the detection of a bright new X-ray source, MAXI\,J1659$-$152, a series of observations
was triggered with almost all currently flying high-energy missions. We report here on 
XMM-Newton, INTEGRAL and RXTE observations during the early phase of the X-ray outburst of this transient black-hole candidate.
We confirm the dipping nature in the X-ray light curves. 
We find that the dips recur on a period of 2.4139$\pm$0.0005\,hrs, and interpret this as the orbital
period of the system. It is thus the shortest period black-hole X-ray binary known to date.
Using the various observables, we derive the properties of the source. The inclination of the accretion disk with respect to the
line of sight is estimated to be 60--75$^{\circ}$. The companion star to the black hole is possibly a M5 dwarf star, with 
a mass and radius of about 0.15\,M$_{\odot}$ and 0.23\,R$_{\odot}$, respectively. The system is rather compact (orbital separation 
is about 1.35\,R$_{\odot}$) and is located at a distance of roughly 7\,kpc. 
In quiescence, MAXI\,J1659$-$152 is expected to be optically faint, about 28\,mag in the V-band.}

\kword{MAXI J1659$-$152 --- black hole --- XMM-Newton, INTEGRAL and RXTE}

\maketitle
\thispagestyle{empty}

\section{Introduction}

\begin{figure*}[t]
\centering
\psbox[xsize=17.5cm]
{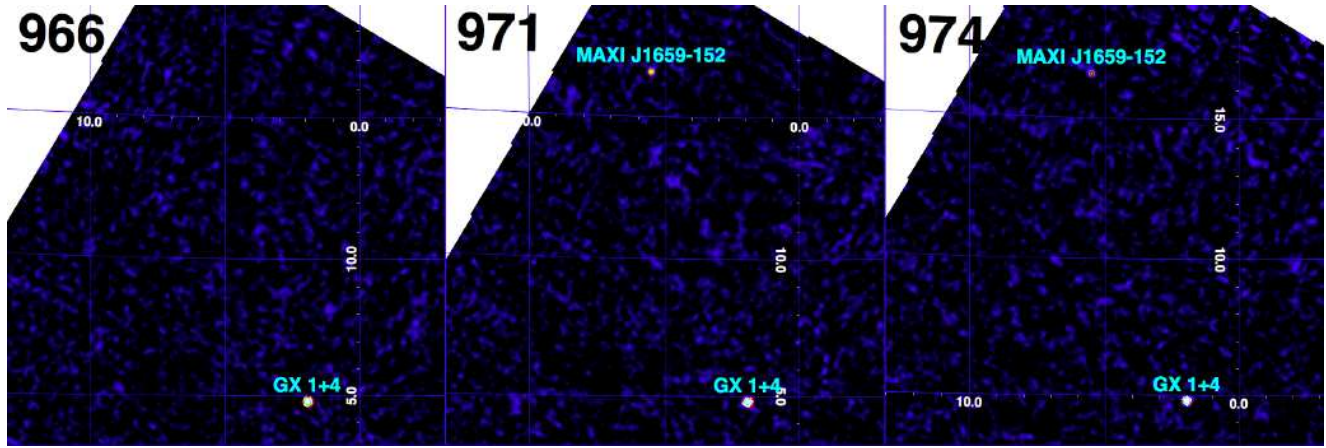}
\caption{Three images taken from observations performed as part of the INTEGRAL Galactic bulge monitoring program
({\tt http://integral.esac.esa.int/BULGE/}). Before the detection of MAXI\,J1659$-$152, in INTEGRAL revolution 966, 2010 September 10/11
({\it Left}), during the first observation after the detection, in revolution 971, September 25/26 ({\it Middle}) and 
during a phase of fading hard X-ray emission, in revolution 974, October 6 ({\it Right}). 
MAXI\,J1659$-$152 is far off-axis, but luckily still within the IBIS/ISGRI field of view, 
so its behaviour was well monitored. Taken from the INTEGRAL Picture of the Month (POM), November 2010 (see {\tt http://www.sciops.esa.int/index.php?project=INTEGRAL\&page=POM}).}
\label{GB}
\end{figure*}

On 2010 September 25 08:05 UT, the Swift/BAT triggered on a source located roughly 17$^{\circ}$ above the Galactic centre 
(see, e.g., Fig.~\ref{GB}).
The source was initially designated as GRB\,100925A (Mangano et al.\ 2010), and monitored with the Swift/XRT. 
Interestingly, the source flux did not decline during the next several hours as is usually the trend with 
GRBs. This unusual behavior and the source location near the Galactic bulge, indicated that it 
might not be a GRB but a new Galactic source (Kahn 2010). 
Later that day, the MAXI\footnote{MAXI stands for Monitoring of All-sky X-ray Image. Note that {\it Maxi} (or {\it Makishi}) was a wandering pirate from Shuri in the 
Ry\=uky\=u Kingdom (present-day Okinawa), see the Soulcalibur Wiki at {\tt http://soulcalibur.wikia.com/wiki/Maxi}.}/GSC team reported the detection of a 
new hard X-ray transient, MAXI\,J1659$-$152, whose position was consistent with GRB\,100925A (Negoro et al.\ 2010).
The Galactic origin was also suggested the next day through a UV-X-ray spectral-energy distribution analysis 
(Xu et al.\ 2010).
Optical spectroscopy by the ESO/VLT X-shooter showed various broad emission lines from the Balmer series of H and
He\,{\sc II}, as well as Ca\,{\sc II} and Na\,{\sc I} absorption from the interstellar medium, all at
redshift zero. The emission lines had double-peaked profiles. These findings suggested the source to be an X-ray 
binary (de Ugarte Postigo et al.\ 2010).
Vovk et al.\ (2010) reported hard X-ray emission from MAXI\,J1659$-$152 on September 27, using
serendipitous INTEGRAL observations at the end of September 25 and the beginning of September 26. 
On September 28, the Swift/XRT team reported frequent dips, possibly due to eclipses by the companion star.
No periodicity could be determined from the available data (Kennea et al.\ 2010a). 
That same day, the fast-timing behaviour observed by the RXTE/PCA 
was shown to be similar to that seen in stellar-mass black-hole transients (Kalamkar et al.\ 2010).
Besides the excellent coverage at high energies, a large multi-wavelength observing campaign has been carried out 
(including optical, submm and radio, see, e.g., van der Horst et al.\ 2010, and references therein).

\section{Observations}

\begin{figure*}[t]
\centering
\psbox[xsize=17.5cm,rotate=l]
{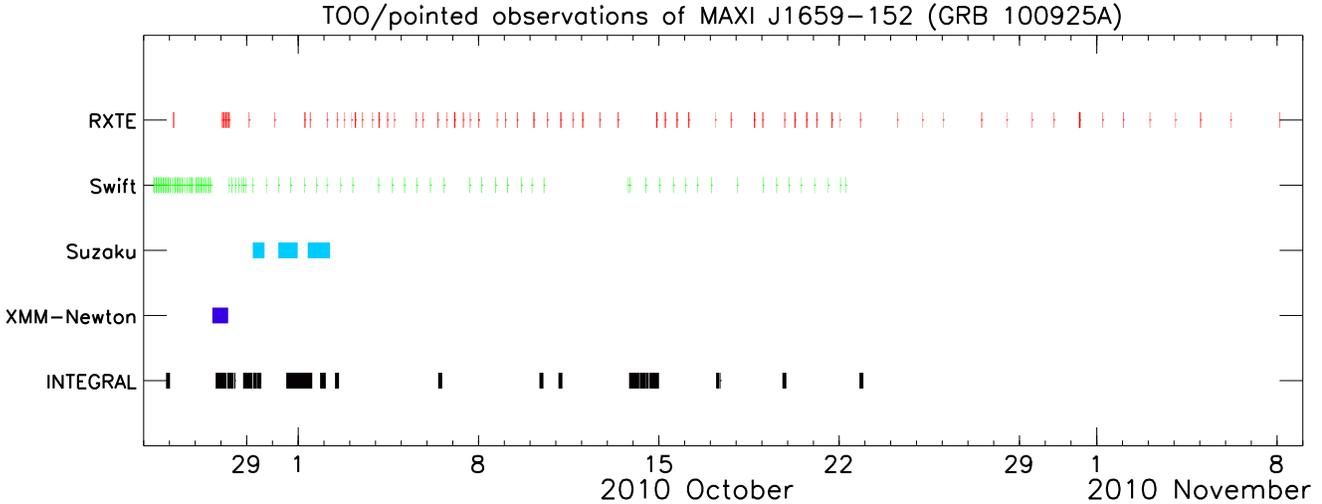}
\caption{After the detection on 2010 September 25 of MAXI\,J1659$-$152 (aka GRB\,100925A), the source was followed by many of the currently flying
high-energy observatories, as shown in the time line above. The serendipitous observations from the INTEGRAL Galactic bulge monitoring program are indicated by the smaller black bars at the bottom.}
\label{multi}
\end{figure*}

After the new source was confirmed, several dedicated observing programs were triggered (see Fig.~\ref{multi}).
Our team was involved in various Target of Opportunity (ToO) observations taken with 
RXTE (Bradt et al.\ 1993), XMM-Newton (Jansen et al.\ 2001) and INTEGRAL (Winkler et al.\ 2003).

XMM-Newton observed MAXI\,J1659$-$152 from UT September 27 15:58 to September 28 06:41 for a total exposure time of 53\,ks
(see \mbox{Kuulkers} et al.\ 2010a; Observation Id: 0656780601).
INTEGRAL observed mostly simultaneously with XMM-Newton, from UT September 27 19:04 to September 29 09:04
(see \mbox{Kuulkers} et al.\ 2010b). Two further INTEGRAL ToO observations were performed 
from September 30 to October 2 and from October 13 to 15.
The INTEGRAL Galactic bulge program (\mbox{Kuulkers} et al.\ 2007) covered MAXI\,J1659$-$152 every few days (see Vovk et al.\ 2010, \mbox{Kuulkers} et al.\ 2010c), albeit at a large off-axis angle (see Fig.~\ref{GB}).
RXTE monitored the source frequently (i.e., more than daily) throughout the outburst (see Kalamkar et al.\ 2010, 2011,
Belloni et al.\ 2010a,b, Shaposhnikov \&\ Yamaoka 2010, Mu\~noz-Darias et al.\ 2010, 2011).
We used 32 PCA observations from September 28 to October 11; the total exposure was about 72.5\,ks.

We used SAS version 10.0.2 together with the latest calibration files to analyze the XMM-Newton data. 
The EPIC-MOS cameras were not used during the observation in order to allocate their telemetry
to the EPIC-pn camera and to avoid Full Scientific Buffer in the latter. The EPIC-pn was used in timing mode. 
Standard data reduction procedures (SAS tasks {\tt epproc} and {\tt rgsproc}) were used to obtain EPIC-pn 
and RGS calibrated event files. We used {\tt epfast} on the event files to correct for a charge
transfer inefficiency (CTI) effect seen in the EPIC-pn timing mode when high count rates
are present. The EPIC-pn exposure was clearly affected by pile-up. To remove this effect, 
we used {\tt epatplot}, which utilizes the relative ratios of single- and
double-pixel events which deviate from standard values in case of significant pile-up,
as a diagnostic tool.
The EPIC-pn and RGS1 time series were corrected and background substracted 
using {\tt epiclccorr} and {\tt rgslccorr}, respectively.

For the INTEGRAL analysis of the JEM-X and IBIS/ISGRI data we used the {\sc OSA9} software.
Light curves were created, using the standard procedures.

For the RXTE/PCA data we used the latest available {\sc FTOOLS} analysis suite.
We produced light curves from PCU2 with 16-s bins over the full PCA range and over the 2--4.5\,keV range.
Since the average count rate of the source varies over the outburst, for each observation 
interval (corresponding to an RXTE orbit), we subtracted the mean count rate. 

\section{X-ray light curves}

\begin{figure}[t]
\centering
\psbox[xsize=8.5cm,rotate=r]
{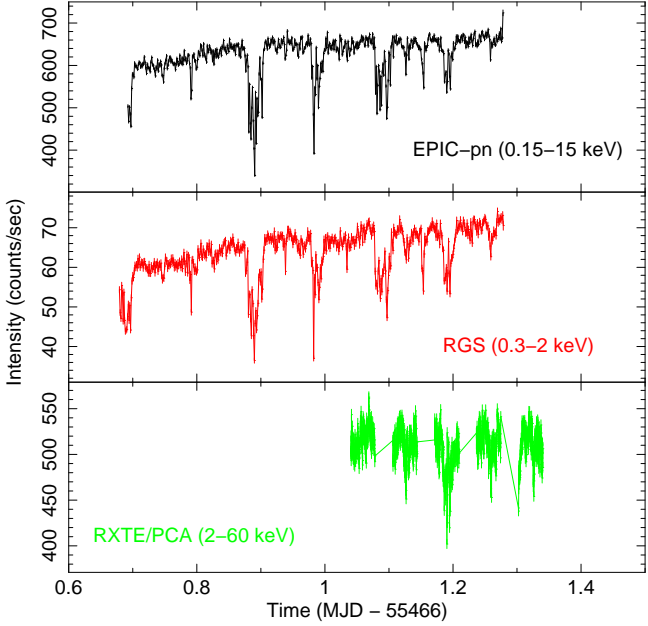}
\caption{XMM-Newton/EPIC-pn (0.15--15\,keV), XMM-Newton/RGS (0.3--2\,keV) and RXTE/PCA (2--60\,keV) light curves on
2010 September 27--28. The time resolution is 100\,s and 16\,s, respectively, for the XMM-Newton and RXTE data.
The data points are connected for clarity.} 
\label{lc}
\end{figure}

The EPIC-pn and RGS1 light curves (Fig.~\ref{lc}, top panels) clearly reveal variations in the 
light curve, which are similar to those reported from the Swift/XRT observations. They recur every 
$\sim$2.4\,hrs, and show irregular structure which lasts between about 5 and 40\,min. 
Occasionally, intermittent, shallower, dips are also seen (see \mbox{Kuulkers} et al.\ 2010d). The 
dips are at most about 50\%\ of the average out-of-dip flux. 
The out-of-dip count rates steadily increase during the course of the observations,
indicating the source X-ray flux was still rising after the Swift/XRT observations (Kennea et al.\ 2010a).

The PCA light curves show the same dip-like structures (e.g., Fig.~\ref{lc}, bottom) 
with a typical length of about 30\,min.
The presence of dips in the JEM-X data is less clear, mainly because of its lower
sensitivity. 

We performed a phase dispersion minimization search (the light curves are highly non-sinusoidal)
on the EPIC-pn and RGS, as well as the renormalized PCA data, using 20 bins with a phase bin width of 0.05, 
over the period range 0.25-12\,hrs. 
The error on the period found was computed by constructing 1000 synthesized data sets.
These data sets were obtained by distributing each data point around 
its observed value, by an amount given by its error bar multiplied by a number output by a Gaussian 
random-number generator with zero mean and unit variance. The measured standard deviation of the 
positions of the deepest troughs in the resulting periodograms was taken as the error.
The period search on the PCA data shows a clear excess in the $\chi^2$ at a period of 2 hours and 
24.83 minutes (with an error of 2\,s), 
corresponding to a deep minimum in the folded light curve centered on T$_0= {\rm MJD}\,55465.17784$
(see also Belloni et al.\ 2010a).
The search using the EPIC-pn and RGS data resulted in a period of 2\,hrs and 24.7\,min
(with an error of 15\,s), consistent with the PCA period. 
Lomb-Scargle analysis of the Swift/XRT data revealed a similar period of 2\,hrs and 25.2\,min
(with an error of 5.4\,min; Kennea et al.\ 2010b).
We also note that optical variations up to 0.1\,mag were reported by Kuroda et al.\ (2010) at the above period 
(2\,hrs and 24.95\,min, with an error of 1\,s).

\section{Discussion}

\begin{figure}[t]
\centering
\psbox[xsize=8.5cm,rotate=r]
{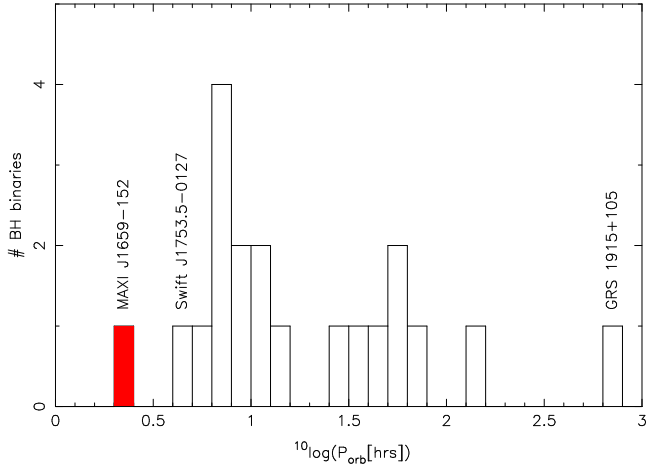}
\caption{Orbital period distribution of all black-hole (candidate) binaries with known orbital periods (all values, except for MAXI\,J1659$-$152, from Ritter \&\ Kolb 2003).
The system with the shortest orbital period is MAXI\,J1659$-$152 (red rectangle), while that with the longest orbital period is GRS\,1915+105. Also indicated is the system Swift\,J1753.5$-$0127.}
\label{porb_dist}
\end{figure}

By analogy with other low-mass X-ray binaries, we interpret the dip structures as X-ray absorption dips 
(rather than eclipses) occurring every orbital period. The dip morphology is very similar to, e.g.,
the black-hole candidate binary 4U\,1755$-$33, which has an orbital period of 4.4\,hr (White et al.\ 1984).
In Fig.~\ref{porb_dist}, we show the orbital period distribution of all black-hole (candidate) binaries 
(Ritter \&\ Kolb 2003). The fastest revolving binary was 
Swift\,J1753.5$-$0127 (Zurita et al.\ 2008: 3.2443\,hrs). It is interesting to note that
Swift\,J1753.5$-$0127 and MAXI J1659$-$152 are both found at high Galactic latitudes (12.2$^{\circ}$ for the former),
as well as two other short-period black-hole binaries (XTE\,J1118+480 and GRO\,J0422+32; see Zurita et al.\ 2008).
As suggested by \mbox{Kuulkers} et al.\ (2010d), if the compact object in 
MAXI\,J1659$-$152 is indeed a black hole (Kalamkar et al.\ 2010, 2011, Belloni et al.\ 2010b, 
Shaposhnikov \&\ Yamaoka 2010, Mu\~noz-Darias et al.\ 2010, 2011), its $\simeq$2.4\,hrs period is to 
our knowledge the shortest among the currently known sample. 

Back-of-the-envelope calculations using various existing empirical relationships can be done
to get an idea of the dimensions of the source. To derive uncertainties on the resulting values, we 
randomly distributed the observed values or equation parameters around their values using a Gaussian distribution with
width equal to their errors, and assumed that all values and parameters are independent.
The spread in the resulting values was used as the uncertainty.

The presence of dips and the absence of eclipses suggest an orbital inclination of about 60--75 degrees 
(see Frank et al.\ 1987).\footnote{Note that Frank et al.\ (1987) adopted a general mass ratio of $q=0.5$; however, their 
results are not very sensitive to $q$.}
Assuming the companion star to the black-hole to be close to the main-sequence, and that it fills its 
Roche-lobe, we can derive estimates of its mass ($M_2$) and its radius ($R_2$), as well as its spectral type,
just using our knowledge of the orbital period, $P_{\rm orb}$ (Smith \&\ Dhillon 1998).
We use the linear relationships for cataclysmic variable (CV) secondaries
(eqs.\ 9 and 12 from Smith \&\ Dhillon 1998), and
derive $M_2\simeq 0.15$\,M$_{\odot}$ and $R_2\simeq 0.23$\,$R_{\odot}$, with an estimated uncertainty of 0.05\,M$_{\odot}$
and 0.02\,$R_{\odot}$, respectively, and a spectral type of about M5 
(must be a dwarf star for such a small star).
Note, however, that low-mass X-ray binary secondaries tend to be a bit smaller and less
massive than CV secondaries at similar $P_{\rm orb}$ (Smith \&\ Dhillon 1998).

Using the commonly accepted fact that a stellar mass black hole has a mass of $M_{\rm BH}\geq 3$\,M$_{\odot}$, we
find a mass ratio of $q=M_2/M_{\rm BH}\leq 0.05$.
Again assuming the companion is Roche-lobe filling, we can use the relation between the orbital separation, $a$, 
and $q$ (Eggleton 1983), to get $a\geq 1.35$\,R$_{\odot}$, i.e., the binary's size is of the order of the Sun.
Assuming the black hole has indeed a mass of 3\,M$_{\odot}$, the duration of the ingress and egress of the 
dips (on the order of 100\,s) and the duration of the dip activity (up to 40\,min) can give a handle on the
extent of the object being absorbed and the absorber itself, respectively (see, e.g., \mbox{Kuulkers} et al.\ 1998, and
references therein). Following \mbox{Kuulkers} et al.\ (1998), we find that the upper limit on the extent of the 
object being absorbed is 0.05\,R$_{\odot}$ or 0.18\,R$_{\odot}$, if the absorbing medium corotates with the binary frame
or corotates with matter in the accretion disk, respectively; the size of the absorbing medium is estimated to be
1.2\,R$_{\odot}$ and 4.3\,R$_{\odot}$, respectively. For larger black-hole masses the estimated sizes are bigger.
Given the size of the object being absorbed we speculate that this
is some kind of (accretion disk) corona surrounding the black hole. The shallow depth of the dips are consistent
with this: part of the corona stays always visible, and hence up to half of the X-ray light is absorbed. 
The estimated sizes of the absorbing medium
are large, i.e., $\sim$1--3 times the orbital separation. This is very unlikely, so we suggest the absorbing medium
to be spread over the outer part of the accretion disk, possibly along the accretion disk rim.

At maximum during outbursts, the optical brightness is dominated by emission from the accretion disk. 
Assuming that all of the optical flux in quiescence comes from the secondary, we can use eq.~5 of Shahbaz \&\ \mbox{Kuulkers} (1998)
to derive an estimate of the absolute disk brightness, which is then only a function of $P_{\rm orb}$.
This leads to a rough estimate of M$_{\rm V,disk}=1.0\pm 0.8$\,mag. 
The observed optical magnitude during outburst maximum was ${\rm V}_{\rm max}\simeq 16.5$ (see Russell et al.\ 2010). 
Using the relation between the hydrogen column density, $N_{\rm H}$, 
and interstellar reddening, $A_{\rm V}$ (Predehl \&\ Schmitt 1995), 
with  $N_{\rm H}=2\times 10^{21}$\,cm$^{-2}$ from the Swift/XRT observations (Kennea et al. 2010a), we derive
we derive $A_{\rm V}\simeq 1.1$ (see also D'Avanzo et al.\ 2010). 
From the distance modulus (see eq.\ 10 of Shahbaz \&\ \mbox{Kuulkers} 1998) we finally infer that the distance to 
MAXI\,J1659$-$152 is about 7\,kpc, with an estimated uncertainty of 3\,kpc. 

Finally, the relation between the outburst amplitude ($\Delta {\rm V} = {\rm V}_{\rm min} - {\rm V}_{\rm max}$) 
and $P_{\rm orb}$ for low-mass X-ray binary transients (Shahbaz \&\ \mbox{Kuulkers} 1998) gives 
$\Delta {\rm V}=11.5\pm 0.8$, and thus the expected V-magnitude in quiescence, V$_{\rm min}=27.9\pm 0.8$.
If true, this is quite faint, and one may therefore not be able to find an optical counterpart in quiescence.
Note, however that Kong et al.\ (2010) reported the Pan-STARRS 1 (PS1) 3Pi sky-survey detection of the source in 
quiescence with about 22.4 in the r-band (AB magnitude). This is brighter compared to our V-band estimate, which cannot be
explained by reddening alone. Possibly the Shahbaz \&\ \mbox{Kuulkers} relation breaks down at short orbital periods, unless the detected source happens to be a foreground star.
We note that, if the outburst amplitude is somewhat lower, the distance estimate becomes smaller. E.g., if $\Delta {\rm V}$ is 1\,mag smaller,
the distance estimate reduces to 4.6\,kpc, with an uncertainty of 2\,kpc.
The maximum observed 2--10\,keV flux during MAXI\,J1659$-$152's outburst ($\simeq$9$\times$10$^{-9}$\,erg\,cm$^{-2}$\,s$^{-1}$; Kennea et al.\ 2010b)
translates to a maximum 2--10\,keV luminosity of roughly 5$\times$10$^{37}$\,erg\,s$^{-1}$, for a distance of 7\,kpc.
This value seems to be a bit faint for a black-hole transient X-ray binary (see, e.g., Dunn et al.\ 2010).
We suggest the low luminosity to be related to the compactness of the system.

\vspace{1pc}
\noindent {\it Acknowledgements:} 
Partly based on observations with INTEGRAL, an ESA project with instruments and science data centre funded by ESA member states (especially the PI countries: Denmark, France, Germany, Italy, Switzerland, Spain), Poland and with the participation of Russia and the USA, and on observations obtained with XMM-Newton, an ESA science mission with instruments and contributions directly funded by ESA Member States and NASA.
We especially thank the XMM-Newton and INTEGRAL Science Operations Centres 
for their prompt scheduling of the Target of Opportunity observations,
5 and 9\,hrs, respectively, between trigger and start of observation on September 27!
We also would like to thank the RXTE Team for their scheduling of the many monitoring observations. 

\section*{References}

\re
Belloni T.M. et al. 2010a ATel., 2926

\re
Belloni T. et al. 2010b ATel., 2927

\re
Bradt H.V. et al. 1993 A\&AS., 97, 355

\re
D'Avanzo P. et al. 2010 ATel., 2900

\re
de Ugarte Postigo A. et al. 2010 GCN Circ., 11307

\re
Dunn R.J.H. et al. 2010 MNRAS., 403, 61

\re
Eggleton P.P. 1983 ApJ., 268, 368

\re
Frank J. et al. 1987 A\&A., 178, 137

\re
Jansen F. et al. 2001 A\&A., 365, L1

\re
Kahn D.A. 2010 GCN Circ., 11299

\re
Kalamkar M. et al. 2010 ATel., 2881

\re
Kalamkar M. et al. 2011 ApJL., submitted [arXiv:1012.4330v1]

\re
Kennea J.A. et al. 2010a ATel., 2877

\re
Kennea J.A. et al. 2010b, Talk presented at the 4th international MAXI Workshop {\em The First Year of MAXI: Monitoring variable X-ray sources},
2010 Nov 30 -- Dec 2, Tokyo, Japan

\re
Kong A.K.H. et al. 2010 ATel., 2976

\re
Kuroda D. et al. 2010, Poster presented at the 4th international MAXI Workshop {\em The First Year of MAXI: Monitoring variable X-ray sources},
2010 Nov 30 -- Dec 2, Tokyo, Japan

\re
Kuulkers E. et al. 1998 ApJ., 494, 753

\re
\mbox{Kuulkers} E. et al. 2007 A\&A., 466, 595

\re
\mbox{Kuulkers} E. et al. 2010a ATel., 2887

\re
\mbox{Kuulkers} E. et al. 2010b ATel., 2888

\re
\mbox{Kuulkers} E. et al. 2010c ATel., 2890

\re
\mbox{Kuulkers} E. et al. 2010d ATel., 2912

\re
Mangano V. et al. 2010 GCN Circ., 11296

\re
Mu\~noz-Darias T. et al. 2010 ATel., 2999

\re
Mu\~noz-Darias T. et al. 2011 MNRAS., submitted

\re
Negoro H. et al. 2010 ATel., 2873

\re
Predehl P. \&\ Schmitt J.H.M.M. 1995 A\&A., 293, 889

\re
Ritter H. \&\ Kolb U. 2003 A\&A., 404, 301 (update RKcat7.14, 2010)

\re
Russell D.M. et al. 2010 ATel., 2884

\re
Shaposhnikov N. \&\ Yamaoka K. 2010 ATel., 2951

\re
Shahbaz T. \&\ \mbox{Kuulkers} E. 1998 MNRAS., 295, L1

\re
Smith D.A. \&\ Dhillon V.S. 1998, MNRAS., 301, 767

\re
van der Horst A.J. et al. 2010 ATel., 2918

\re
Vovk I. et al. 2010 ATel., 2875

\re
White N.E. et al. 1984 ApJ., 283, L9

\re
Winkler C. et al. 2003 A\&A., 411, L1

\re
Xu D. et al. 2010 GCN Circ., 11303

\re
Zurita C. et al. 2008 ApJ., 681, 1458

%\re
%Author B.C. et al. 1997 PASJ., 1111, 1111

\label{last}

\end{document}